\documentclass[12pt,amsbsy,epsf]{article}
\usepackage{graphicx}
\usepackage{amssymb}
\addtocounter{page}{0} \tolerance=5000

\newcommand{\beq}{\begin{equation}}
\newcommand{\eeq}{\end{equation}}
\newcommand{\bJ}{\mbox{${\bf J}$}}
\newcommand{\va}{\mbox{${\bf a}$}}

\newcommand{\al}{\mbox{${\alpha}$}}

\newcommand{\ka}{\mbox{${\kappa}$}}
\newcommand{\om}{\mbox{${\omega}$}}

\begin{document}

\begin{titlepage}
\vspace{1cm}

\begin{center}
{\large \bf Nonthermal radiation of rotating black holes}
\end{center}

\begin{center}
R.V. Korkin\footnote{korkin@vxinpz.inp.nsk.su}, I.B.
Khriplovich\footnote{khriplovich@inp.nsk.su}
\end{center}
\begin{center}
Budker Institute of Nuclear Physics\\
630090 Novosibirsk, Russia,\\
and Novosibirsk University
\end{center}

\bigskip

\begin{abstract}
Nonthermal radiation of a Kerr black hole is considered as
tunneling of created particles through an effective Dirac gap. In
the leading semiclassical approximation this approach is
applicable to bosons as well. Our semiclassical results for
photons and gravitons do not contradict those obtained
previously. For neutrinos the result of our accurate quantum
mechanical calculation is about two times larger than the
previous one.
\end{abstract}

\vspace{7cm}

\end{titlepage}

\section{Introduction}

The amplification of an electromagnetic wave at the reflection
from a rotating black hole, so called superradiation, was
predicted by Zel'dovich \cite{zel}, and also by Misner \cite{mis}.
Then it was studied in detail by Starobinsky and Churilov
\cite{sta} for electromagnetic and gravitational waves (see also
\cite{bek}). It looks rather obvious that if the amplification of
a wave at the reflection is possible, then its generation by a
rotating black hole is possible as well. Indeed, direct
calculation by Page \cite{pag} has demonstrated that the
discussed, nonthermal radiation does exist, and not only for
bosons, photons and gravitons, but for neutrinos as well. The last
result looks rather mysterious since for fermions there is no
superradiation.

In the present work these processes are considered from another
point of view: as tunneling of quanta being created through the
Dirac gap. Certainly, this approach by itself can be valid for
fermions only. It is clear however that in the leading
semiclassical approximation the production of fermions and bosons
is described by same, up to the statistical weight, relationships.

Let us note that in the recent work by Calogeracos and Volovik
\cite{cal} (we found out about it after the present paper had been
written) an analogous mechanism was considered for the description
of the friction experienced by a body rotating in superfluid
liquid at $T=0$: the quantum tunneling of quasiparticles to the
region where their energy in the rotating frame is negative.

Our semiclassical results for photons and gravitons are in a
reasonable qualitative agreement with the calculation \cite{pag}.
One cannot expect here a quantitative coincidence since for the
most essential partial waves the action inside the barrier exceeds
unity not so much if any. As to neutrinos, the imaginary part of
the action for them, at the total angular momenta of real
importance, is considerably smaller then unity. Therefore, for
spin $s=1/2$ we have performed a complete numerical quantum
mechanical calculation of the effect. Here our result exceeds
that presented in \cite{pag} by a factor close to two.
Unfortunately, the lack of details in \cite{pag} does not allow
us to elucidate the origin of the discrepancy.

\section{Scalar field}

We will start with a problem of a methodological rather than
direct physical interest, with the radiation of scalar massless
particles by a rotating black hole.

The semiclassical solution of the problem is started from the
Hamilton-Jacobi equations for the motion of a massless particle
in the Kerr field (see, for instance, \cite{ll}):

\begin{equation}\label{sr}
\left(\frac{\partial{S_{r}(r)}}{\partial{r}}\right)^2=
-\frac{\ka^2}{\Delta}+\frac{[(r^2+a^2)\varepsilon-a\,l_{z}]^2}{\Delta^2}\,,
\end{equation}

\begin{equation}\label{sth}
\left(\frac{\partial{S_{\theta}(\theta)}}{\partial{\theta}}\right)^2=
 \ka^2
 -\left(a\,\varepsilon\,\sin{\theta}-\frac{l_{z}}{\sin{\theta}}\right)^2.
\end{equation}
Here $S_{r}(r)$ and $S_{\theta}(\theta)$ are the radial and
angular actions, respectively;
\[
\Delta=r^2-r_g r +a^2; \quad r_g=2kM;
\]
$\va= \bJ/M$ is the angular momentum of the black hole in the
units of its mass $M$; $\varepsilon$ is the particle energy;
$l_{z}$ is the projection of the particle angular momentum onto
$\va$ (the velocity of light $c$ is put to unity everywhere).

As to the constant $\ka^2$ of the separation of variables, in the
spherically symmetric limit $a\rightarrow\infty$ it is equal to
the particle angular momentum squared ${\bf l}^2$, or to $l(l+1)$
in quantum mechanics (the Planck constant $\hbar$ is also put to
unity everywhere). The influence of the black hole rotation, i.e.
of the finite $a$, upon $\ka^2$ is taken into account by means of
the perturbation theory applied to equ. (\ref{sth}). The result is
\cite{sta}
\begin{equation}\label{ka}
\ka^2=l(l+1)-2\omega\,\alpha\,l_{z}+
\frac{2}{3}\om^2\,\al^2\left[1+\frac{3l^2_{z}-l(l+1)}{(2l-1)(2l+3)}\right].
\end{equation}
Here and below we use the dimensionless variables $\om
=\varepsilon k M, \quad x = r/kM, \quad \al = a/kM$. Let us
recall that in the semiclassical approximation the substitution
\[l(l+1)\rightarrow (l+1/2)^2\]
should be made.

Besides, in the exact quantum mechanical problem, under the
reduction of the radial wave equation to the canonical form
\[
R'' + p^2(r) R =0,
\]
in the expression for $p^2(r)$ additional (as compared to the
right-hand side of equ. (\ref{sr})) nonclassical terms arise,
which should be, strictly speaking, included for $l\simeq 1$.
However, for the simplicity sake we neglect here and below these
nonclassical corrections to $p^2(r)$, which does not influence
qualitatively the results obtained.

The dependence of the classically inaccessible region, where the
radial momentum squared $p^2$ is negative, on the distance $x$ is
presented for different angular momenta in Figs. 1, 2. At the
horizon the gap vanishes \cite{der}. For $r \rightarrow\infty$
the boundaries  of the classically inaccessible region behave as
$\pm (l+1/2)/r$. In other words, the centrifugal term for
massless particles plays in a sense the role of the mass squared.
Let us note that for $l>1$ both branches of the equation
$p^2(r)=0$ fall down, but for $l=1$ one branch near the horizon
grows up, and the second one falls down. Thus, the radiation
mechanism consists in tunneling, in going out of particles from
the dashed region to the infinity.

One should note the analogy between the emission of charged
particles by a charged black hole and the effect discussed. In
the first case the radiation is due to the Coulomb repulsion, and
in the case considered here it is due to the repulsive
interaction between the angular momenta of the particle and black
hole \cite{unr}.

\begin{figure}
\centering \includegraphics[height=10cm]{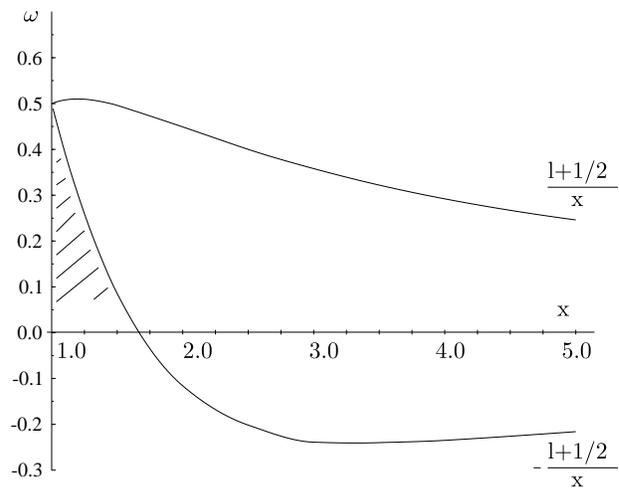} \caption{The
energy gap for $l=1$.}
\end{figure}

\begin{figure}
\centering \includegraphics[height=10cm]{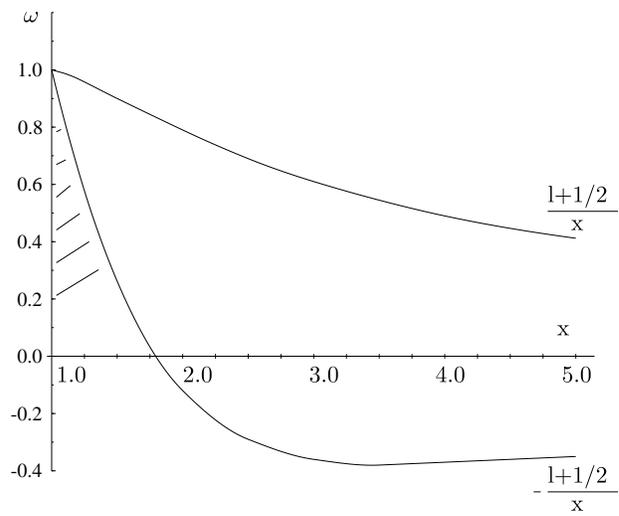} \caption{The
energy gap for $l=2$.}
\end{figure}

The action inside the barrier for the radial equ. (\ref{sr}) is:
\begin{equation}
 |S_{r}|=\int dx\,
\sqrt{\frac{k^2}{(x-1)^2}-\frac{[\omega\,(x^2+1)-l_{z}]^2}{(x-1)^4}}\,;
\end{equation}
the integral is taken between two turning points. For simplicity
sake we confine for the time being to the case of an extremal
black hole, $a=kM$. Let us note that due to a singular dependence
of $p$ on $x$, the action inside the barrier does not vanish at
$l>1$ even for the maximum energy $\om =l_z/2$. So much the more,
it stays finite at $l=1$ (compare Figs. 1 and 2).

The repulsive interaction is proportional to the projection $l_z$
of the particle angular momentum and enters the tunneling
probability through the exponent, but the barrier depends on the
orbital angular momentum $l$ itself. Therefore, clearly the main
contribution to the effect will be given by particles with $l_z$,
close to $l$. The numerical calculation demonstrates that the
contribution of the states with $l_z \neq l$ can be neglected at
all. Besides, since the action inside the barrier decreases with
the growing energy, the main contribution to the effect is given
by the particles with energy close to maximum.

Unfortunately, even for an extremal black hole an analytical
calculation of the action inside the barrier cannot be carried out
to the end. Therefore, to obtain a qualitative idea of the effect,
we will use a simplified expression for $\ka^2$:
\beq\label{kap}
\ka^2=l^2+l-2\omega\,l+\omega^2.
\eeq
(The results of a more accurate numerical calculation with
expression (\ref{ka}) will be presented below.) In this
approximation one can obtain a simple analytical formula for
action inside the barrier for all angular momenta but $l=1$. Let
us assume that $\omega= (1-\delta)\,l/2$ with $\delta\ll 1$; just
this range of energies contributes most of all into the radiation.
Then, the turning points of interest to us, which are situated to
the right of the horizon, are:
\beq\label{x12}
x_{1,2}=1+\frac{2\delta}{2\pm \sqrt{1+ 4/l}}.
\eeq

Now one finds easily
\beq\label{san}
|S_{an}|=\,\frac{\pi\,l}{2}\left(2-\sqrt{3-\frac{4}{l}}\right).
\eeq
One can see from this equation that the term $l$ (following
$l^2$) in formula (\ref{kap}) is quite essential even for large
angular momenta: it generates the terms $4/l$ in formulae
(\ref{x12}) and (\ref{san}), thus enhancing $|S|$ for $l\gg 1$ by
$\pi/\sqrt{3}$. Correspondingly, the transmission factor
$D=\exp(-2|S|)$ gets about $40$ times smaller. Let us note that
even the transition in $\ka^2$ from $l(l+1)$ to $(l+1/2)^2$ makes
the effect considerably smaller for $l$ comparable to unity; but
this suppression dies out for large angular momenta.

It follows from formula (\ref{san}) that the action inside the
barrier is large, it increases monotonically with the growth of
$l$, starting with $|S|=\pi$ for $l=2$. As to $l=1$, one can see
by comparing Fig. 1 with Fig. 2 that here the barrier is wider
than for $l=2$, and therefore the action should be larger. Indeed,
the numerical calculation of the action inside the barrier $|S|$
with $\ka^2$ given by formula (\ref{ka}) confirms these estimates.
Its results are presented in Table 1 where for the comparison sake
we present also the analytical estimates $|S_{an}|$ with formula
(\ref{kap}). By the way, this comparison demonstrates that the
approximate analytical formula (\ref{kap}) works
\begin{table}
[h]
\begin{center}
\begin{tabular}{|c|c|c|c|}\hline
    &        &        &      \\
$l$ & $1$    & $2$    & $3$      \\
    &        &        &  \\

 \hline
             &        &        &  \\
$|S|$        & $3.45$ & $3.15$ & $3.33$  \\
    &        &        & \\
\hline
    &        &        & \\
$|S_{an}|$ &        & $3.14$ & $3.34$  \\
    &        &        & \\ \hline
\end{tabular}
\end{center}
\begin{center}
Table 1: Action inside the barrier for scalar particles.
\end{center}
\end{table}

\noindent very well. The numbers presented in the table refer to
an extremal black hole and maximum energy of emitted particles. It
is clear however that the transition to nonextremal black holes,
lower energies, and larger $l$ will result in the growth of the
action inside the barrier. Since here it proves to be always
considerably larger than unity, the use of the semiclassical
approximation inside the barrier is quite reasonable.

Let us check now whether it applies to the left of the barrier.
The corresponding condition is of the usual form:
\beq\label{scc}
\frac{d}{d\,x}\frac{1}{p(x)} \ll 1.
\eeq
In other words, the minimum size of the initial wave packet to
the left of the barrier should not exceed the distance from the
horizon to the turning point. Near the horizon one can neglect in
the expression for the momentum $p(x)$ the term related to the
centrifugal barrier, so that
\[
\frac{d}{d\,x}\frac{1}{p(x)}\approx
\frac{d}{d\,x}\frac{(x-1)^2}{\omega\,(x^2+\alpha^2)-\alpha\,l}.
\]
One can see easily that for not so large $l$, which are of
importance in our case, this expression is comparable to unity
and condition (\ref{scc}) does not hold. Nevertheless, despite of
this circumstance and of the neglect of the nonclassical
corrections to $p^2(r)$ mentioned above, the results of the
semiclassical calculation, presented below, are correct
qualitatively.

Let us come back to the calculation of the radiation intensity.
The radial current density of free particles in the energy
interval $d\varepsilon$ is at $r \to \infty$
\begin{equation}
j_{\,r}(\varepsilon,l)d\varepsilon = \sum
\frac{d^3p}{(2\,\pi)^3}\,\frac{\partial\varepsilon}{\partial
p_{\,r}} =\sum_{l,l_z} \frac{2\,\pi \, d p_{\,r} }{(2\,\pi)^3\,
r^2 }\,\frac{\partial\varepsilon}{\partial p_{\,r}}\,.
\end{equation}
Indeed, the initial summation here reduces to the integration over
the azimuth angle of the vector ${\bf l}$, which gives $2\pi$,
and to the account for the contributions of all projections $l_z$
of the orbital angular momentum. As pointed out already, in our
case it is sufficient to take into account only one of them,
$l_z=l$. By means of the identity
\[
\frac{\partial\varepsilon}{\partial p_{\,r}}\,dp_{\,r} =
d\varepsilon,
\]
we obtain in the result that the total flux of free particles at
$r \to \infty$ is
\begin{equation}
4\pi r^2 j_{\,r}(\varepsilon,l) = 4\pi r^2 \sum_{l_z}
\frac{2\,\pi}{(2\,\pi)^3 r^2} \to \frac{1}{\pi}.
\end{equation}

One can easily see that in our problem the total flux of radiated
particles differs from the last expression by the barrier
penetration factor only. Thus, in our semiclassical approximation
we obtain for the loss of mass by a black hole in the unit time
the following expression:
\begin{equation}\label{M8}
\frac{dM}{dt}=\frac{1}{\pi}\sum_{l=1}^\infty
\int\limits^{\varepsilon_{max}}_{0}\varepsilon
\exp(-2\,|S(\varepsilon,l)|)\,d\varepsilon\,.
\end{equation}
Here the maximum energy of radiated quanta is
\beq\label{em}
\varepsilon_{max}= \,\frac{a l}{r_h^2+a^2};
\eeq
$r_h = km + \sqrt{k^2 M^2 - a^2}$ is the radius of the horizon of
a Kerr black hole. The analogous expression for the loss of the
angular momentum is
\begin{equation}\label{M9}
\frac{dJ}{dt}=\frac{1}{\pi}\sum_{l=1}^\infty
\int\limits^{\varepsilon_{max}}_{0} l
\exp(-2\,|S(\varepsilon,l)|)\,d\varepsilon\,.
\end{equation}

The results of the numerical calculation with formulae (\ref{M8})
and (\ref{M9}) of the loss by a black hole of its mass and angular
momentum for different values of the rotation parameter $\alpha$
are presented in Table 2. We present here and below, for spinning
particles, results of calculations only for sufficiently rapid
rotation, $\alpha \approx 1$. The point is that with further
decrease of $\alpha$, not only the thermal radiation grows
rapidly, but the effect discussed falls down even more rapidly.
For smaller $\alpha$ this effect becomes much smaller than the
thermal one, and thus its consideration there does not make much
sense.
\begin{table}
[h]
\begin{center}
\begin{tabular}{|c|c|c|} \hline
         &          &               \\
$\alpha$ &$|\,dM/dt|$ & $|\,dJ/dt| $     \\
         &          &               \\
\hline
         &          &               \\
$0.999$ & $2.6$ & $6.4$  \\
         &          &               \\
\hline
         &          &               \\
$0.9$ &  $0.19$ & $0.77$ \\
         &          &               \\
\hline
\end{tabular}
\end{center}
\begin{center}
Table 2: Loss of mass (in the units of $10^{-3}\,\pi M^2$) and
angular momentum (in the units of $10^{-3}\,\pi M$), due to the
radiation of scalar particles.
\end{center}
\end{table}

As one can see from Table 2, the rate of loss of the angular
momentum is higher, in the comparable units, than the rate of
loss of the mass. In fact, it follows immediately from expression
(\ref{em}). Already from this expression one can see that even
for the maximum possible energy the ratio of the corresponding
numbers is 2:1. Real ratios are even larger. Hence an important
conclusion follows: extremal black holes do not exist. Even if an
extremal hole is formed somehow, in the process of radiation it
looses the extremality immediately.

\section{Radiation of photons and gravitons}

The investigation of the radiation of real particles we start
with the electromagnetic field. Photon has two modes of opposite
parity: the so called electric mode, with $l=j \pm 1$, and
magnetic one, with $l=j$~\cite{blp}. It follows from the duality
invariance that the radiation intensities for these two modes are
equal. Thus one can confine to the solution of the problem for
the magnetic mode, and then just double the result.

One can demonstrate that the situation with the gravitational
waves is analogous. Again, there are two modes which, due to a
special duality, contribute equally to the radiation, and for one
of these modes $l=j$.

Obviously, for a mode with $l=j$ the radial equation in the
semiclassical approximation is the same as for the scalar field,
but with different value of $\ka^2$. It can be demonstrated also
starting from the so called Teukolsky equation \cite{teu}
(neglecting again nonclassical corrections to $p^2(r)$). The
corresponding eigenvalues of the angular equation for particles
of spin $s$, found again in the perturbation theory, are
\cite{sta}:
\[
\ka^2=j(j+1)+\frac{1}{4}-2\alpha\,\omega\,j_{z}-
\frac{2\alpha\,\omega\,j_{z}\,s}{j(j+1)}+\alpha^2\omega^2\left\{
\frac{2}{3}\left[1+\frac{3j^2_{z}-j(j+1)}{(2j-1)(2j+3)}\right]\right.
\]
\begin{equation}\label{kasp}
\left. -\,\frac{2s^2}{j(j+1)} \frac{3j^2_{z}-j(j+1)}{(2j-1)(2j+3)}
+2\,s^2\left[\frac{(j^2-s^2)(j^2-j^2_{z})}{j^2(2j-1)(2j+1)}-
\frac{((j+1)^2-j^2_{z})((j+1)^2-s^2)}{(j+1)^3(2j+1)(2j+3))}\right]\right\}.
\end{equation}
We have included into this expression the term $1/4$, necessary
for correct semiclassical description. Let us note that, as
follows from the consideration of the helicity of a massless
particle, the restriction $j\geq s$ holds. Correspondingly, for a
photon $j\geq 1$, for a graviton $j\geq 2$. As well as in the
scalar case, the main contribution into the radiation is given by
the states with the maximum projection of the angular momentum,
$j_z=j$.

Let us discuss first of all whether the semiclassical
approximation is applicable here. As to the situation to the left
of the barrier, it does not differ qualitatively from the scalar
case. The situation inside the barrier is different. As one can
see from equation (\ref{kasp}), the presence of spin makes $\ka^2$
smaller, and correspondingly, makes smaller the centrifugal
repulsion. In result, both the barrier and the action inside it
decrease. This qualitative argument is confirmed by a numerical
calculation of $|S|$ for photons and gravitons with the maximum
projection of the angular momentum $j_{z}=j$ and maximum energy
for the case of an extremal black hole (see Table 3).
\begin{table}
[h]
\begin{center}
\begin{tabular}{|c|c|c|c|c|} \hline
 \multicolumn{1}{|c|}{}& \multicolumn{2}{|c|}{s=1}&
\multicolumn{2}{|c|}{s=2} \\  \hline &&&&\\ $j$ &$1$&$2$ &$2$&$3$ \\
&&&&\\ \hline &&&&\\ $|S|$ & $1.84$ & $2.17$&$1.0$&$1.7$  \\ &&&&\\
\hline
\end{tabular}
\end{center}
\begin{center}
Table 3: Action inside the barrier for photons and gravitons.
\end{center}
\end{table}
Therefore in the present case, one should expect that the
accuracy of semiclassical results is lower than in the scalar
case.

The semiclassical formulae for electromagnetic and gravitational
radiation differ formally from the corresponding scalar ones
(\ref{M8}) and (\ref{M9}) by extra factor 2 only, which reflects
the existence of two modes. The results of this calculation are
presented in Table 4. In it we indicate in brackets for
comparison the results of the complete quantum mechanical
calculation \cite{pag} which takes into account as well the
thermal radiation.
\begin{table}
[h]
\begin{center}
\begin{tabular}{|c|c|c|c|c|} \hline
\multicolumn{1}{|c|}{}& \multicolumn{2}{|c|}{s=1}&
\multicolumn{2}{|c|}{s=2} \\ \hline
&&&&\\
$ \alpha  $ & $|dM/dt|$ & $|dJ/dt|$ & $|dM/dt| $  & $|dJ/dt|$     \\
 &&&&\\
\hline &&&&\\ $0.999$ & $16.5 (9.6)$ & $39 (24)$ &$66 (228)
$&$148(549) $ \\  &&&&\\
\hline &&&&\\ $0.9$ &  $0.72 (2.26)$ & $2.8 (8.2)$ &$0.58 (12.9)$&$2 (48)$ \\
&&&&\\
\hline
\end{tabular}
\end{center}
\begin{center}
Table 4: Loss of mass (in the units of $10^{-3}\,\pi M^2$) and
angular momentum (in the units of $10^{-3}\,\pi M$), due to the
radiation of photons and gravitons.
\end{center}
\end{table}
It is clear from Table 4 that even for $\alpha=0.999$, when the
thermal radiation is negligibly small, our semiclassical
calculation agrees with the complete one qualitatively only. It
is only natural if one recalls that the semiclassical action in
the present problem exceeds unity not so much, if any. This
explanation is supported by the fact that for photon, where $|S|$
is considerably larger (see Table 3), the agreement of the
semiclassical calculation with the complete one is considerably
better.

\section{Radiation of neutrino}

Let us consider at last the radiation by a rotating black hole of
neutrinos, massless particles of spin $1/2$. The wave function of
a 2-component neutrino is written as (see, for instance
\cite{pag}):
\begin{equation}
\psi= \exp(-i\varepsilon\,t + i j_z\phi) \left(\begin{array}{c}
R_{1}\,S_{1}
\\R_{2}\,S_{2}
\end{array}\right).
\end{equation}
It is essential that the wave equations for neutrino in the Kerr
metrics allow for the separation of variables as well \cite{teu}.
The radial equations in dimensionless variables are:
\[
\frac{d\,R_{1}}{d\,x}-i\,\frac{\omega(x^2+\alpha^2)-j_{z}\,\alpha}
{\Delta}\,R_{1}=\frac{\kappa}{\sqrt{\Delta}}\,R_{2},
\]
\begin{equation}\label{ra}
\frac{d\,R_{2}}{d\,x}+i\,
\frac{\omega(x^2+\alpha^2)-j_{z}\,\alpha}
{\Delta}\,R_{2}=\frac{\kappa}{\sqrt{\Delta}}\,R_{1}.
\end{equation}
The angular equations are:
\[
\frac{d\,S_{1}}{d\,\theta}+
\left(\omega\,\alpha\,\sin{\theta}-\frac{j_{z}}{\sin{\theta}}\right)
\,S_{1}=\kappa\,S_{2},
\]
\begin{equation}
\frac{d\,S_{2}}{d\,\theta}-
\left(\omega\,\alpha\,\sin{\theta}-\frac{j_{z}}{\sin{\theta}}\right)
\,S_{2}=-\kappa\,S_{1}.
\end{equation}
For $\ka^2$ the same formula (\ref{kasp}) takes place, but now of
course with $s=1/2$. As well as for bosons, it is sufficient
practically to consider states with $j_z=j$.

It is essential that $R_1$ corresponds at infinity, for $x \to
\infty$, and at the horizon, for $x \to 1$, to the wave running to
the right, and $R_2$ corresponds for $x \to \infty$ and for $x \to
1$ to the wave running to the left. (For this classification, it
is convenient to use the so called ``tortoise'' coordinate
$\xi(x)$; for $x \to \infty \quad \xi \approx x \to +\infty$, for
$x \to 1 \quad \xi \approx \ln (x-1) \to -\infty$.) It is quite
natural that here the radial current density is
\[
j_r = |R_{1}|^{2}-|R_{2}|^{2}.
\]

We are interested in the probability of penetrating the barrier
for the state which is an outgoing wave at infinity. For a
neutrino or antineutrino such a state has a fixed helicity, but
has no definite parity. Meanwhile, the potential barrier depends
in our problem, roughly speaking, on the orbital angular momentum,
and therefore is much more transparent for the states of
$l=j-1/2$, than for the states of $l=j+1/2$. (These states of
given $l$ have definite parity, and are superpositions of neutrino
and antineutrino.) Moreover, at $l=j-1/2$ for small $j$, which
give the main contribution to the radiation, the action either has
no imaginary part at all, or its imaginary part is small, so that
our above approach is inapplicable. Therefore, we will solve
numerically the exact problem of the neutrino radiation.

Technically, it is convenient to find the reflection coefficient
$R$ in the problem of the neutrino scattering off a black hole,
and then use the obvious relationship for the transmission
coefficient~$D$:
\[
D=1-R.
\]
Here the expressions for the loss of mass and angular momentum by
a black hole are:
\begin{equation}
\frac{dM}{dt}=\frac{1}{\pi}\sum_{j=1/2}^\infty
\int\limits^{\varepsilon_{max}}_{0}\varepsilon D(\varepsilon, j)
\,d\varepsilon\,;
\end{equation}
\begin{equation}
\frac{dJ}{dt}=\frac{1}{\pi}\sum_{j=1/2}^\infty
\int\limits^{\varepsilon_{max}}_{0} j D(\varepsilon, j)
\,d\varepsilon\,;
\end{equation}
\beq
\varepsilon_{max}= \,\frac{a j}{r_h^2+a^2}.
\eeq
The results, obtained by numerical solution of the system of
radial equations (\ref{ra}), are presented in Table 5. In brackets
we present the results of \cite{pag}, which include the Hawking
radiation contribution. For a black hole close to extremal one, at
$\alpha = 0.99$, where the thermal radiation is practically
absent, our results are about twice as large as previous ones.
\begin{table}
[h]
\begin{center}
\begin{tabular}{|c|c|c|} \hline
         &          &               \\
$\alpha$ &$|\,dM/dt|$ & $|\,dJ/dt| $     \\
         &          &               \\
\hline
         &          &               \\
$0.99$   & $4.4 \quad (2.1) $ & $11 \quad (5.65)$  \\
         &          &               \\
\hline
         &          &               \\
$0.9$ &  $0.7 \quad (1)$ & $2.7 \quad (3.25)$ \\
&          &               \\
\hline
\end{tabular}
\end{center}
\begin{center}
Table 5: Loss of mass (in the units of $10^{-3}\,\pi M^2$) and
angular momentum (in the units of $10^{-3}\,\pi M$), due to the
radiation of neutrinos.
\end{center}
\end{table}

\begin{center} *** \end{center}
We are grateful to G.E. Volovik who has attracted our attention
the article \cite{cal}. The present work was supported in part by
the Russian Foundation for Basic Research through Grant No.
01-02-16898, through Grant No. 00-15-96811 for Leading Scientific
Schools, by the Ministry of Education Grant No. E00-3.3-148, and
by the Federal Program Integration-2001.

\end{document}